# Effect of ligand substitution on the exchange interactions in {Mn$_{12}$}-type single-molecule magnets

Danil W. Boukhvalov[1,2], Viatcheslav V. Dobrovitski[3], Paul Kögerler[3,4], Mohammad Al-Saqer[3], Mikhail I. Katsnelson[2], Alexander I. Lichtenstein[5], and Bruce N. Harmon[3]

[1]Computational Materials Science Center, National Institute for Materials Science, 1-2-1 Sengen, Tsukuba, Ibaraki 305-0047, Japan

[2] Radboud University Nijmegen, Institute for Molecules and Materials, Heyendaalseweg 135, 6525 AJ Nijmegen, the Netherlands

[3]Ames Laboratory, Iowa State University, Ames, IA 50011

[4]Institute of Inorganic Chemistry, RWTH Aachen University, 52074 Aachen, Germany

[5]Institute of Theoretical Physics, University of Hamburg, 20355 Hamburg, Germany

E-mail: D.Bukhvalov@science.ru.nl

*We investigate how the ligand substitution affects the intra-molecular spin exchange interactions, studying a prototypal family of single-molecule magnets comprising dodecanuclear cluster molecules [Mn$^{III}_8$Mn$^{IV}_4$O$_{12}$(COOR)$_{16}$]. We identify a simple scheme based on accumulated Pauling electronegativity numbers (a.e.n.) of the carboxylate ligand groups (R). The redistribution of the electron density, controlled by a.e.n. of a ligand, changes the degree of hybridization between 3d electrons of manganese and 2p electrons of oxygen atoms, thus changing the exchange interactions. This scheme, despite its conceptual simplicity, provides a strong correlation with the exchange energies associated with carboxylate bridges, and is confirmed by the electronic structure calculations taking into account the Coulomb correlations in magnetic molecules.*

# 1. Introduction

Single-molecule magnets (SMM) attract much attention from chemists and physicists. Crystals of SMM constitute ensembles of identical, weakly interacting nanosized magnetic entities, and therefore provide a unique opportunity for studying nanomagnets quantum behavior, such as quantum tunneling of magnetization, the role of topological phases, etc [1]. Moreover, the recently discovered phenomenon of spin-sensitive tunneling through individual molecules makes SMM potentially important for future applications in nanoelectronics [2].

$[Mn_{12}O_{12}(COOR)_{16}](H_2O)_4CH_3COOH$ (with R = $CH_3$), here abbreviated as $Mn_{12}$, is one of the first synthesized [3] and best known SMM. Earlier investigations [4] have shown that the magnetization curves in external magnetic fields, as well as the temperature dependence of the magnetic susceptibility, are different for different carboxylate ligands (R = $CH_3$ and $C_6H_5$). X-ray and photoemission spectroscopy results [5, 6] show that $Mn_{12}$ molecules with different R have different electronic structure. In another SMM system, {$Mn_4$} [7], the changes in chemical composition of the ligands provide even more changes in the magnetic properties of the system. During the last decade, many types of {$Mn_{12}$} clusters with different ligands have been synthesized, and their magnetic properties are strongly R-dependent [4, 8-19]. Variations in the magnetic anisotropy, in the effective barrier of magnetization reversal, and in the magnetic relaxation, caused by substitution of the ligands have been discussed in these papers. Dependence of magnetic properties on the solvents and preparation routes has also been considered in several works [20]. The ligand substitution may be useful for optimization of the magnetic properties of SMM needed for future applications. Detailed theoretical understanding of the relation between the chemical structure of ligands, their electronic structure, and the magnetism may provide important guidelines for fine-tuning the properties of SMMs.

Here we explore a different approach that aids understanding these relations. We consider dependence of the intra-molecular exchange interactions on the properties of ligands. This issue has been addressed in much less detail than the anisotropy, although it clarifies interesting and important interplay between the electronic properties of ligands and exchange couplings. Such interplay can be used to tune the energy separation from the ground-state spin multiplet and the excited multiplets with other value of the total spin. This allows controlling the temperature dependence of magnetic susceptibility and the mixing between different spin multiplets due to Dzyaloshinsky-Moriya interactions [21]. It also allows tuning the rate of the phonon-assisted transitions between different spin multiplets, thus affecting the SMM's spin relaxation properties. Moreover, such tuned SMMs might be useful for applications in single-molecule spintronics, where the excited multiplets determine voltage dependence of the transmission through SMM.

## 2. Structural properties of $Mn_{12}$

The {$Mn_{12}$} molecule contains a "ferrimagnetic" metal-oxide core $Mn_{12}O_{12}$ (see Figure 1a) with opposite directions of the magnetic moments on $Mn^{3+}$ (Mn2 and Mn3 on Figure 1a) and $Mn^{4+}$ (Mn1 on Figure 1a) sites. Every Mn1 ion is connected to other Mn1, Mn2 and two of Mn3 ions by the oxygen bridges (single bridge with Mn3, and double bridge with Mn2), and by the carboxylate bridges which are connected to the ligands R (the ligands under consideration are listed in Table 1). The four Mn3 ions are coordinated by

H$_2$O; in some structures (e.g. {Mn$_{12}$} with R = CH$_3$) all four Mn3 are coordinated by one H$_2$O, while in {Mn$_{12}$} with R = C$_6$F$_5$, only two Mn3 are coordinated by two H$_2$O. This difference in water coordination does not influence essentially the electronic structure and the exchange interactions, but turns out to be very important for the redox properties of {Mn$_{12}$} [19, 22]. A scheme of the exchange interactions is displayed in Figure 1b. Both experiment [23] and theory [24] suggest that for {Mn$_{12}$} with R = CH$_3$ all exchange couplings are antiferromagnetic, with J1 being close to J2, and both being stronger than J3 and J4. Variation of ligands does not alter significantly the geometry of the inner core of the molecule: the distances Mn-Mn, Mn-O-Mn, and the angles Mn-O-Mn change by less than 2%. However, change of the ligand R may change the space group symmetry not only for arrangement of the molecules in a crystal, but also for the molecule's point group symmetry. In the case of high symmetry (I-4), all four Mn$^{4+}$ denoted as Mn1 are equivalent. When the symmetry is low (P-1, P2$_1$/c, P2$_1$/n, etc.) all Mn$^{3+}$ ions are not equivalent, but their magnetic moments and oxygen environments are very similar, so that these four Mn$^{3+}$ ions can still be treated as equivalent. In a similar way, other non-equivalent but sufficiently similar Mn ions (and their exchange couplings) in the case of low-symmetry molecules are described as belonging to the same classes. Below, we consider only the point symmetries of individual molecules, which primarily govern the electronic structure and the magnetic properties of SMM. Correspondingly, all {Mn$_{12}$} molecules below can be classified as having high or low symmetry. Along with changing the symmetry of the molecule, the change in ligands may significantly alter the electronic structure of the SMM by redistributing the electron density, therefore altering the magnetic exchange interactions between the Mn spin centers.

## 3. Computational Method

To study these effects in detail, we use a Local Density Approximation (LDA) of the density functional theory, taking into account the on-site Coulomb repulsion [25] (LDA+U approach). The account of the local Coulomb interaction is crucial for an adequate description of the transition metal-oxide systems in general, and for $\{Mn_{12}\}$ SMMs in particular. The value of the Coulomb parameter U for $\{Mn_{12}\}$ is 4 eV as has been determined earlier based on experimental and theoretical studies [6, 23] of $\{Mn_{12}\}$. For our electronic structure calculation we use the LMTO-ASA [25] method implemented in the Stuttgart TB-47 code. This approach was successfully adopted by us earlier for studying $\{Mn_{12}\}$ [6, 24], $\{Mn_6\}$ [27], $\{Mn_4\}$ [7], $\{V_{15}As_6\}$ [28], $\{V_{12}As_8\}$ [29], and $\{Fe_8\}$ [30] spin clusters. The spin-polarized LDA+U calculations taking into account the spin-orbit interactions are computationally expensive and could be used only for simple systems [31]. In the present work we consider only the electronic structure and exchange interactions in $\{Mn_{12}\}$ systems.

We have performed calculations for several existing types of $\{Mn_{12}\}$, and for some model structures which have not been synthesized yet, but are close to existing systems. The changes in the projected densities of states (DOS) caused by variation of the ligands are summarized in Figs. 2, 3. They suggest that the distributions of the electron density in the ligands and in the core of the $\{Mn_{12}\}$ molecules are strongly correlated. These correlations exist because the d-electrons are not very strongly localized on Mn sites but rather they are noticeably hybridized with oxygens. Correspondingly, the exchange couplings are not local characteristics of the manganese-oxide bridges, but are properties of the molecule as a whole. Using the calculated exchange parameters, we can construct a corresponding spin Hamiltonian (assuming isotropic Heisenberg exchange), and find its eigenvalues and eigenfunctions [24]. For all ligands considered here, we find that the ground state corresponds to the total molecular spin $S = 10$, in agreement with the experimental data.

## 4. Accumulated electronegativity of ligands

In order to quantitatively characterize how different ligands redistribute the electron density, and therefore how they affect the exchange interactions, we examined a large number of relevant chemical constants, and found that more appropriate for this purpose is the bond dissociation energy (BDE) for R-H analogues of the R-COO$^-$ carboxylate groups. This constant characterizes the strength of the covalent bonds between hydrogen and R, and is valid for description of bonds between R and COO- groups in {Mn$_{12}$}. Unfortunately, the experimental results for BDE in R-COOH systems are available only for few groups; the reported values of BDE for the compounds used in the ligands in the {Mn$_{12}$} systems are shown in Table 1. Since for a large number of chemical species used for the ligand substitution in {Mn$_{12}$} the BDE data are not available, we introduce an alternative quantitative descriptor, the accumulated electronegativity numbers (a.e.n.) of ligands, and use it instead of BDE for quantitative characterization of the ligands. We define the accumulated electronegativity of a ligand as a sum of the electronegativities (e.n.) of constituent elements, for instance a.e.n.(-CH$_3$) = e.n.(C) + 3 e.n.(H), where e.n. of the individual atoms are standard and well known. In the same way, we can calculate a.e.n. for different R: a.e.n.(-CH$_2$CH$_2$Cl) = 2 e.n.(C) + 4 e.n.(H) + e.n.(Cl), a.e.n.(-

C4H4SCH3) = 4 e.n.(C) + 4 e.n.(H) + e.n.(S), where the e.n. of the CH3 tail is not taken into account since the methyl group is located very far from the carboxylate bridge, and does not influence its electronic structure noticeably. Of course, this way of calculating the a.e.n. of the ligands has many drawbacks. First, there are many possible ways of defining the electronegativity of the atoms: we use the Pauling electronegativity scale. This choice is sufficient for the purposes of our study, since the differences between various e.n. scales are relatively small and not important for the approximate account given here. Second, different parts of a ligand affect the electronic structure in a different way, so that the e.n. of different atoms should be included in the sum with different weights (depending on the atom's position, its neighbors, etc.). Such a fine-tuned approach requires much work and is underway now. However, since we study a narrower group of ligands, it is expected that the simplified definition above is sufficient for our purposes.

## 5. Results and Discussion

In Fig. 4, we show the connection between the magnitude of the intramolecular exchange interactions and the e.n. of ligands for both low- and high-symmetry {Mn$_{12}$} molecules. This connection can be qualitatively understood as follows. Fig. 2 demonstrates strong hybridization between the *3d* orbitals of Mn atoms from the molecule's inner region and the *2p* orbitals of O from the carboxylate bridges: the main peaks in the density of the *3d* states are also visible in the *2p* DOS. At the same time, oxygens from the carboxylate bridges are connected through carbon atoms with R, thus connecting the ligands and the inner region of {Mn$_{12}$}. In Figure 3, one can see that the replacement of the ligands influences strongly the degree of *p-d* hybridization. With an increasing a.e.n. of the ligands the main peaks of Mn *3d* and O *2p* bands move in different directions and become closer to each other. For R = $CH_3$ this distance is about 1.5 eV (the peak of Mn *3d* DOS is above the peak of O *2p* DOS), while for R = $C_6H_5$ the peak of O *2p* DOS already lies higher than the Mn *3d* DOS peak by about 0.2 eV. The calculated centers of gravity for Mn *3d* and O *2p* DOS move with the change of a.e.n. in a similar way: Mn *3d* DOS moves towards lower energies, and O *2p* DOS moves towards higher energies. Figs. 2 and 3 show dramatic differences in the electronic structure between R = $C_6H_5$ and other ligands. Along with the large a.e.n. of $C_6H_5$, this difference is also caused by the reduced symmetry (P-1 for R = $C_6H_5$ and I-4 for the others). The symmetry strongly affects the exchange parameters, see Fig. 4. Further increase in a.e.n. by substitution of hydrogens with fluorine atoms (R=$C_6F_5$) leads to insignificant shifts of Mn *3d* and O *2p* bands towards each other (see decrease of distance between peaks a and c on Fig. 3c, d) as it was discussed for R=$C_6H_5$; this results in further enhancement of ferromagnetic exchanges and decrease of antiferromagnetic exchanges (see Fig. 4). Currently, the majority of synthesized {$Mn_{12}$} molecules have the I-4 point symmetry (except for the {$Mn_{12}$} with charged solvent [26]). One may speculate that the effect of the lower symmetry is more important than the effect of redistribution of the electron density, but this conjecture requires a separate extensive investigation. Using the calculated exchange interactions, we calculated the total spin of the molecule in its magnetic ground state. Note that for all carboxylate groups we obtained an S = 10 ground state. Therefore, for experimental check of the calculated intra-molecule exchange parameters, the energies of the excited spin states should be measured (e.g., by inelastic neutron scattering) and compared to the calculated ones.

## 6. Conclusions

In summary, we found a pronounced and systematic correlation between chemical structure of the carboxylate ligands and exchange interactions in $\{Mn_{12}\}$ species. Different types of $\{Mn_{12}\}$ compounds exhibit the same spin ground state, but the intramolecular exchange interactions and other magnetic properties can differ dramatically. These results demonstrate that the exchange interactions are the properties of the molecule as a whole, due to delocalization of the *3d* electrons via *p-d* hybridization. The correlations between the chemical structure of the ligands, the symmetry of molecule, and the exchange interactions may be useful for synthesizing new varieties of $\{Mn_{12}\}$ compounds with desirable magnetic properties. The ligands substitution allows tuning the energies of the excited multiplets with different total spin, which affects the spintronic properties of SMMs, the temperature dependence of magnetic susceptibility, and the spin relaxation behavior. The ground-state total spin of the spin cluster can also be modified by further increasing of a.e.n. of ligands, which would change all exchanges from antiferromagnetic to ferromagnetic. Although the concept of a ligand a.e.n. is not yet a rigorous quantitative notion, the crude approximation used above already provides a consistent and systematic model interpretation of electronic structure obtained from detailed calculations. It may be interesting to further refine the definition of the ligand a.e.n., taking into account the properties and mutual arrangement of different groups. This requires more studies, and such a work is underway.

**Acknowledgment** The work is financially supported by Stichting voor Fundamenteel Onderzoek der Materie (FOM), the Netherlands. Work at the Ames Laboratory was supported by the Department of Energy – Basic Energy Sciences under Contract No. DE-AC02-07CH11358.

# References

1. Friedman, J.; Sarachik, M.; Tejada, J.; Zolio, R. *Phys. Rev. Lett.* **1996**, *76*, 3830; Thomas, L.; Lionti, F.; Ballou, R.; Gatteschi, D.; Sessoli, R.; Barbara, B. *Nature (London)* **1996**, *383*, 145; Wernsdorfer, W.; Sessoli, R.; Caneschi, A.; Gatteschi, D.; Cornia, A. Europhys. Lett. **2000**, *50*, 552; Ardavan, A.; Rival, O.; Morton, J. J. L.; Blundell, S. J.; Tyryshkin, A. M.; Timco, G. A.; Winpenny, R. E. P. *Phys. Rev. Lett.* **2008**, *98*, 057201; Bertaina, S.; Gambarelli, S.; Mitra, T.; Tsukerblatt, B.; Müller, A.; Barbara, B. *Nature (London)* **2008**, *453*, 203; Cornia, A.; Sessoli, R.; Sorace, L.; Gatteschi, D.; Barra, A. L. Daiguebonne, C. *Phys. Rev. Lett.* **2002**, *89*, 257201; Hill, S.; Edwards, R. S.; Jones, S. I.; Dalal, N. S.; North, J. M. *Phys. Rev. Lett.* **2003**, *90*, 217204; del Barco, E.; Kent, A. D.; Hill, S.; North, J. M.; Dalal, N. S.; Rumberger, E. M.; Hendrikson, D. N.; Chakov, N.; Christou, G. *J. Low Temp. Phys.* **2005**, *140*, 119.

2. Bolink, J. H.; Capelli, L.; Coronado, E.; Recalde, I. *Adv. Mater.* **2006**, *18*, 920; Heersche, H. B.; de Groot, Z.; Folk, J. A.; van der Zant, H. S. J.; Romeike, C.; Wegewijs, M. R.; Zobbi, L.; Barreca, D.; Tondello, E.; Cornia, A. *Phys. Rev. Lett.* **2006**, *96*, 206801; Mannini, M.; Pineider, F.; Sainctavit, P.; Joly, L.; Fraile-Rodríguez, A.; Cartier dit Arrio, M.-A.; Moulin, C.; Wernsdorfer, W.; Cornia, A.; Gatteschi, D.; Sessoli, R. *Adv. Mater.* **2009**, *21*, 167.

3. Lis T. *Acta Cristalogr. Sect. B* **1980**, *36*, 2042.

4. Sessoli R.; Tsai H.-L.; Schake A.R.; Wang S.; Vincent J.B.; Folting K.; Gatteschi D.; Christou G.; Hendrickson D.N. *J. Am. Chem. Soc.* **1993**, *115*, 1804.

5. Voss, S.; Fonin, M.; Rudiger, U.; Burgert, Groth, U.; Dedkov, Yu. S. *Phys. Rev. B* **2007**, *75*, 045102.

6. del Pennio, U.; De Renzi, V.; Biagi, R.; Corradini, V.; Zobbi, L.; Cornia, A.; Gatteschi, D.; Bondino, F.; Magnano, E.; Zangrando, M.; Zacchigna, M.; Lichtenstein, A.; Boukhvalov, D. W. *Surf. Sci.* **2006**, *600*, 4185.

7. Kampert, E.; Janssen, F. F. B. J.; Boukhvalov, D. W.; Russcher, J. C.; Smits, J. M. M.; de Gelder, R.; de Bruin, B.; Christianen, P. C. M.; Zeitler, U.; Katsnelson, M. I.; Maan, J. C.; Rowan, A. E. *Inorg. Chem.* **2009**, *48*, 11903.

8. Chakov, N. E.; Lee, S.-C.; Harter, A. G.; Kuhus, P.L.; Reyes, A.P.; Hill, S. O.; Dalal, N. S.; Wernsdorfer, W.; Abboud, K. A.; Christou G. *J. Am. Chem. Soc.* **2006**, *128*, 6975.

9.  Tsai, H.-L.; Shiao, H.-A.; Jwo, T.-Y.; Yang, C.-I; Wur, C.-S.; Lee, G.-H. *Polyhedron* **2005**, *24*, 2205.

10. Zhao, H.; Berlinguette, C.P.; Bacsa, J.; Prosvirin, A.V.; Bera, J.K.; Tichy, S. E.; Schelter, E. J.; Dunbar, K. R.; *Inorg. Chem.* **2004**, *43*, 1359.

11. Zobbi, L.; Mannini, M.; Pacchioni, M.; Chastanet, G.; Bonacchi, D.; Zandradi, C.; Biagi, R.; del Pennino, U.; Gatteschi, D.; Cornia, A.; Sessoli, R. *Chemm. Commun.* **2005**, 1640.

12. Park, C.-D.; Rhee, S.W.; Kim, Y.L.; Jeon, W.S.; Jung, D.-Y.; Kim, D.-H.; Do, Y.; Ri, H.-C. *Bull. Korean Chem. Soc.* **2001**, *22*, 453.

13. Yoon, S.W.; Heu, M.; Jeon, W.S.; Jung, D.-Y.; Suh B.J.; Yoon, S. *Phys. Rev. B* **2003**, *67*, 052402.

14. Lim, J. M.; Do, Y.; Kim, J. *Eur. J. Inorg. Chem.* **2006**, 711-717.

15. Soler, M.; Artus, P.; Folting, K.; Huffman, J.C.; Hendrickson, D.N.; Christou, G. *Inorg. Chem.* **2001**, *40*, 4902.

16. Aubin, S. M. J.; Sun, Z.; Guzei, I.A.; Reingold, A.L.; Christou, G.; Hendrickson, D.N. *Chemm. Commun.* **1997**, 2239.

17. Soler, M.; Wernsdorfer, W.; Sun, Z.; Huffman, J.C.; Hendrickson, D.N.; Christou, G. *Chemm. Commun.* **2003**, 2672.

18. Sun, Z.; Ruiz, D.; Rumberger, E.; Incarvito, C. D.; Folting, K.; Reingold, A.L.; Christou, G.; Hendrickson, D.N. *Inorg. Chem.* **1998**, *37*, 4758.

19. Chakov, N. E.; Soler, M.; Wernsdorfer, W.; Abboud, K. A.; Christou, G. *Inorg. Chem.* **1999**, *38*, 5329.

20. Hill, S.; Anderson, N.; Wilson, A.; Takahashi, S.; Chakov, N. E.; Murugesu, M.; North, J. M.; Dalal, N. S.; Christou, G. *J. Appl. Phys.* **2005**, *97*, 10M510; Redler, G.; Lampropoulos, C.; Datta, S.; Koo, C.; Stamatatos, T. C.; Chakov, N. E.; Christou, G.; Hill, S. *Phys. Rev. B* **2009**, *80*, 094408; Domingo, N.; Luis, F.; Nakano, M.; Muntó, M.; Gómes, J.; Chaboy, J.; Ventosa, N.; Campo, J.; Veciana, J.; Ruiz-Molina, D. *Phys. Rev. B* **2009**, *79*, 214404; Carbonera, C.; Luis, F.; Campo, J.; Sánchez-Marcos, J.; Camón, A.; Chaboy, J.; Ruiz-Molina, D.; Imaz, I.; van Slagern, J.; Dengler, S, González, M. *Phys. Rev. B* **2010**, *81*, 014427.

21. Katsnelson, M. I.; Dobrovitski, V. V.; Harmon, B. N. *Phys. Rev. B* **1999**, *59*, 6919; Herzog, S.; Wegewijs, M. R. *Nanotechnology* **2010**, *21*, 274010; Chaboussant, G.; Sieber, A.; Ochsenbein, S.; Güdel, H.-U.; Murrie, M.; Honecker, A.; Fukushima, N. Normand, B. *Phys. Rev. B* **2004**, *70*, 104422.

22. Basler, R.; Sieber, A.; Chaboussant, G.; Güdel, H.-U.; Chakov, N. E.; Soler, M.; Christou, G.; Desmedt, A.; Lechner, R.; *Inorg. Chem.* **2005**, *44*, 649; Soler, M.; Wernsdorfer, W.; Abud, K. A.; Huffman, J. C.; Davidson, E. R.; Hendrickson, D. N.; Christou, G. *J. Am. Chem. Soc.* **2001**, *125*, 3576-3588.

23. Chaboussant G.; Sieber A.: Ochsenbein S.; Güdel H.-U.: Murrie M.; Honecker A.; Fukushima N.; Normand B. *Phys. Rev. B* **2004**, *70*, 104422.

24. Boukhvalov, D. W.; Al-Saqer, M.; Kurmaev, E. Z.; Moewes, A.; Galakhov, V. R.; Finkelstein, L. D.; Chiuzbăian, S.; Neumann, M.; Dobrovitski, V. V.; Katsnelson, M. I.; Lichtenstein, A. I.; Harmon, B. N.; Endo, K.; North, J. M.; Dalal, N. S. *Phys. Rev. B* **2007**, *75*, 014419.

25. Anisimov, V.I.; Aryasetiawan, F.; Lichtenstein, A.I. *J. Phys.: Condens. Matter* **1997**, *9*, 767-808.

26. Gunarson, O.; Jepsen, O.; Andersen, O. K. *Phys. Rev. B* **1983**, *27*, 7144.

27. del Pennino, U.; Corradini, V.; Biagi, R.; De Renzi, V.; Moro, F.; Boukhvalov, D. W.; Panaccione, G.; Hochstrasser, M.; Carbone, C.; Milios, C. J.; Brechin E. K. *Phys. Rev. B* **2008**, *77*, 085419.

28. Boukhvalov, D.W.; Kurmaev, E.Z.; Moews, A.; Yablonskih, M.V.; Chiuzbăian, S.; Finkelstein, L.D.; Neuman, M.; Katsnelson, M.I.; Dobrovitski, V.V.; Lichtenstein, A.I. *J. Electr. Spectr. Relat. Phenom.* **2004**, *137-140*, 735-739.

29. Boukhvalov, D.W.; Kurmaev, E. Z.; Moewes, A.; Zatsepin, D. A.; Cherkashenko, V. M.; Nemnonov, S. N.; Finkelstein, L. D.; Yarmoshenko, Yu. M.; Neumann, M.; Dobrovitski, V. V.; Katsnelson, M. I.; Lichtenstein, A. I.; Harmon, B. N.; Kögerler, P. *Phys. Rev. B* **2003**, *67*, 134408; Boukhvalov D.W.; Dobrovitski V.V.; Katsnelson M.I.; Lichtenstein A.I.; Harmon B.N. *Phys. Rev. B* **2003**, *70*, 0504417.

30. Barbour, A.; Luttrell, R. D.; Choi, J.; Musfeldt, J. L.; Zipse, D.; Dalal, N. S.; Boukhvalov, D. W.; Dobrovitski, V. V.; Katsnelson, M. I.; Lichtenstein, A. I.; Harmon, B. N.; Kögerler, P. *Phys. Rev. B* **2006**, *74*, 014411.

31. Tsivokin, Y. Y.; Korotin, M. A.; Shorikov, A. O.; Anisimov, V. I.; Voloshinskii, A. N.; Lukoyanov, A. V; Koneva, E. S.; Povzner, A. A.; Surin, M. A. *Phys. Rev. B* 2**007**, *76*, 075119; Mazurenko, V. V.; Anisimov, V. V. *Phys. Rev. B* **2005**, *71*, 184434.

**Table I** Chemical structure of ligands (R) with references, experimental spins in ground states (S), internal symmetry group, bond dissociation energy for H-R compound and accumulated electronegativity of ligands (a.e.n., see description in text).

| R | S | Symmetry group | BDE (kJ/mol) | a.e.n. |
|---|---|---|---|---|
| $CH_3$ [8] | 10 | I-4 | 439.0 | 8.80 |
| $CH_2Br$ [13] | 10 | I-4 | | 9.50 |
| $CF_2Cl$ [14] | 10 | I-4 | | 13.50 |
| $CF_3$ [15] | 10 | I-4 | 445.2 | 14.50 |
| $C_6H_4SCH_3$ [16] | 10 | I-4 | | 24.90 |
| $C_6H_4F$ | model | I-4 | | 27.40 |
| $C_6H_2F_3$ | model | I-4 | | 31.20 |
| $C_6F_5$ | model | I-4 | | 35.00 |
| $CH_2CH_2Cl$ [17] | 10 | P-1 | | 16.40 |
| $C_4H_3S$ [19] | 10 | P-1 | | 17.80 |
| $(CHCl_2)_{0.5}(CH_2C(CH_3)_3)_{0.5}$ [20] | 10 | P-1 | | 20.15 |
| $C_6H_5$ [9] | 10 | P-1 | 472.2 | 25.50 |
| $C_6H_4CH_3$ [21] | 10 | P-1 | | 25.90 |
| $CH_2C(CH_3)_3$ [22] | 10 | P-1 | | 30.60 |
| $CH_2C_6H_5$ [23] | 10 | P-1 | | 32.20 |
| $C_6F_5$ [24] | 10 | P-1 | 487.4 | 35.00 |
| $CH_2C_6F_5$ | model | P-1 | | 41.70 |

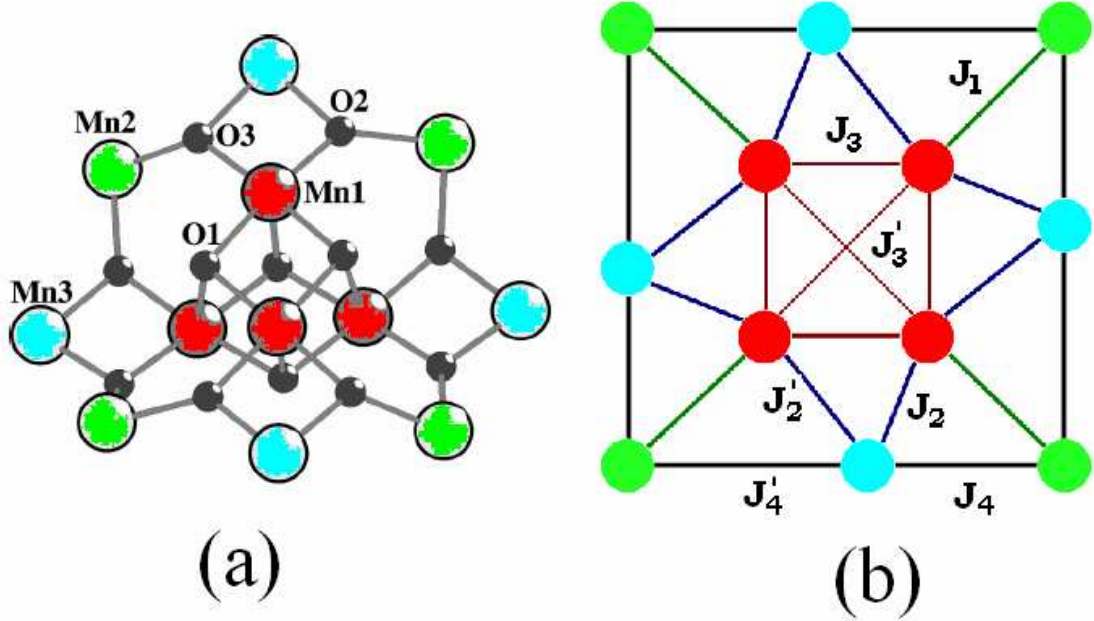

**Figure 1.** Scheme of (a) central part $Mn_{12}O_{12}$, and (b) exchange interactions. Red circles - $Mn^{4+}$, green and blue circles - $Mn^{3+}$, black circles - oxygen.

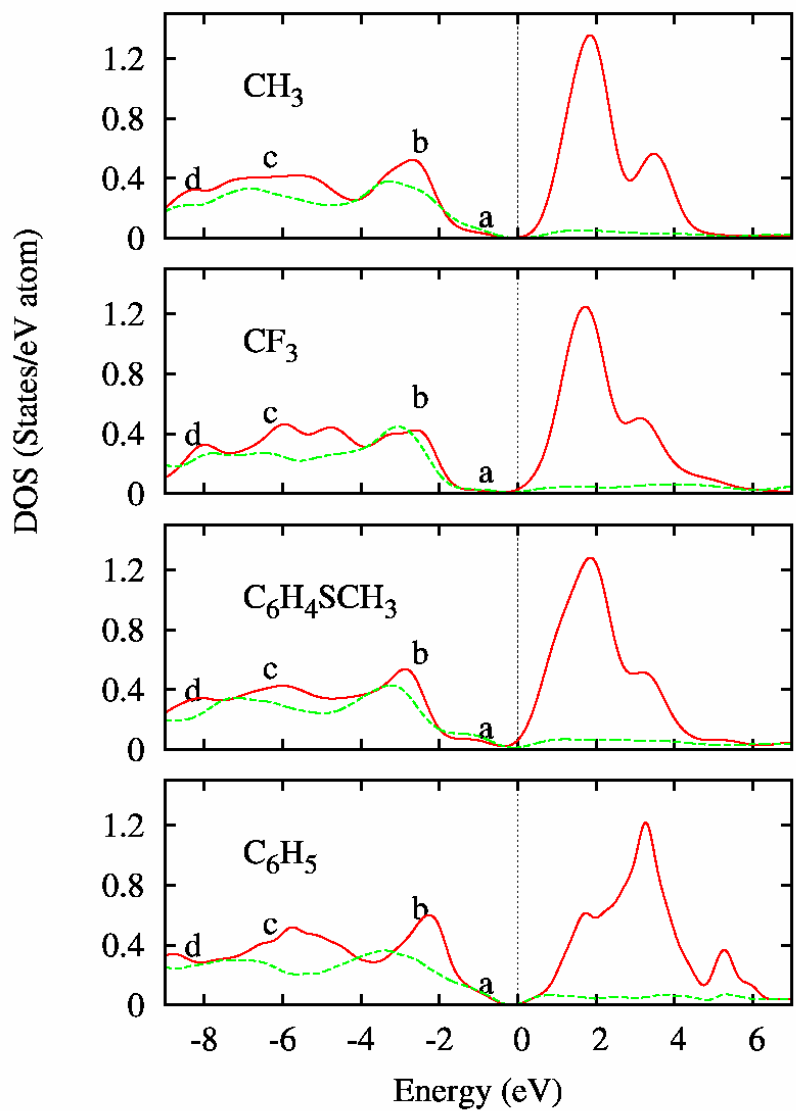

**Figure 2.** Densities of *3d* states for Mn1 (solid red line) and *2p* states for oxygen (dashed green line) from carboxylate bridges for different kinds of {Mn$_{12}$}.

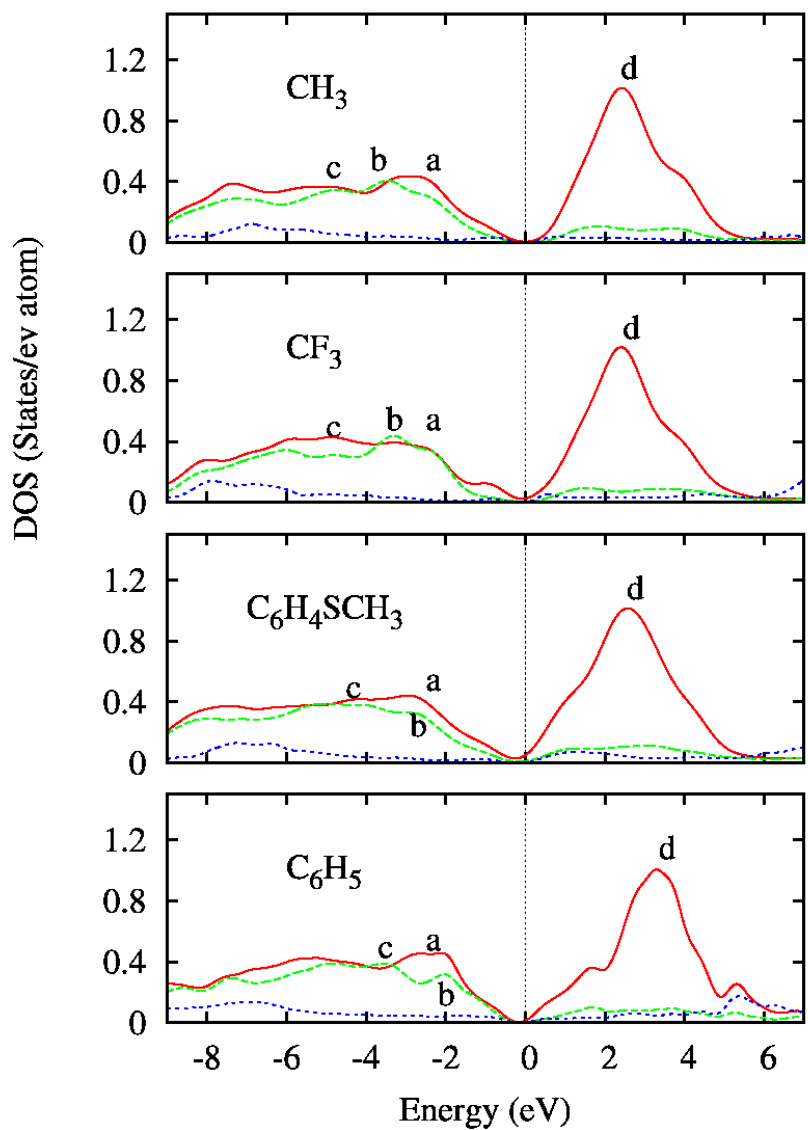

**Figure 3.** Densities of states for *3d* orbitals of all Mn (solid red line), *2p* orbitals of oxygens connected to Mn ions (dashed green line) and *2p* orbitals of carbons from carboxylate bridges (doted blue line).

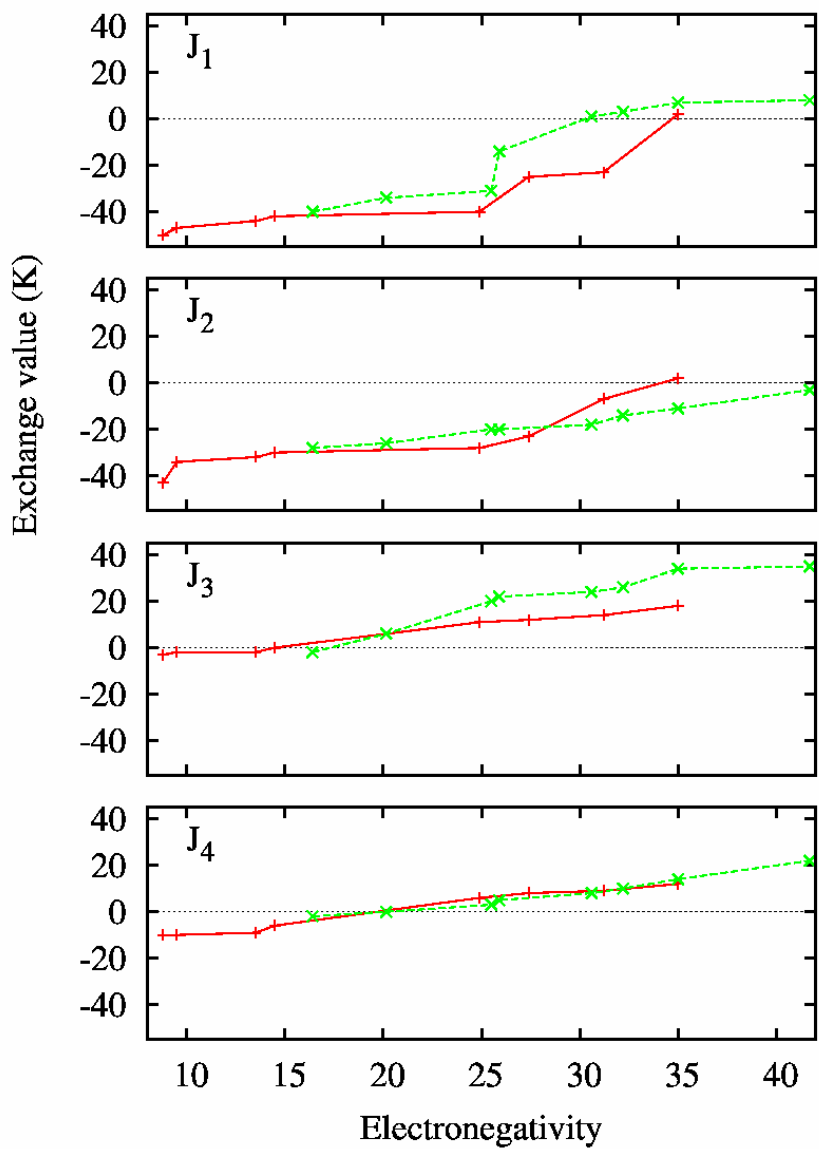

**Figure 4.** Dependence values of exchange interactions for $Mn_{12}$ with high symmetry (solid red line), and low symmetry (dashed green line) of electronegativity of ligands (see description in text).